# Orbital Angular Momentum Multiplexing in Highly Reverberant Environments

Xiaoming Chen, *Senior Member, IEEE*, Wei Xue, Hongyu Shi, *Member, IEEE*, Jianjia Yi, *Member, IEEE*, Wei E. I. Sha, *Senior Member, IEEE*

*Abstract*—Previous studies on orbital angular momentum (OAM) communication mainly considered line-of-sight environments. In this letter, however, it is found that OAM communication with high-order modulation can be achieved in highly reverberant environments by combining the OAM multiplexing with a spatial equalizer. The OAM multiplexing exhibits comparable performance of conventional multiple-input multiple-output (MIMO) system.

*Index Terms*—Multiplexing, orbital angular momentum (OAM), reverberant environment

## I. INTRODUCTION

ORBITAL angular momentum (OAM) has been applied to various applications, such as data multiplexing [1] and radar imaging [2]. Recently, it has been applied to reduce the uncertainty of stochastic measurements [3]. There are different techniques for generating OAM waves. For examples, OAM can be generated using uniform circular arrays (with digital or analogue discrete Fourier transform) [4], [5], spiral phase plates [6], [7], or geometric-phase based metasurfaces [8], [9].

Most research works on OAM communication assume pure line-of-sight (LOS) environment, e.g., [1], [4]-[9]. Only a few of studies have considered OAM transmission in multipath environments. For example, OAM communication has been investigated in an indoor environment [10], where it was found that the OAM communication was sensitive to antenna misalignment and multipath. The multipath effect of specular reflection on OAM communication was studied in [11], where it was shown that higher OAM modes suffer more crosstalk. Although spatial equalizers have been applied to OAM detection in pure LOS environment [12], [13], almost all the OAM detection rely on the orthogonality of OAM modes. Thus, multipath (that destroys the OAM orthogonality) has detrimental effects on the conventional OAM communication.

As pointed out in [14], OAM-based multiplexing does not offer a new degree of freedom. In essence, it also uses the spatial domain like the conventional multiple-input multiple-output (MIMO) system. The transformation from orthogonal plane waves to orthogonal OAM modes does not increase the scattering channels or communication capacities. In fact, OAM multiplexing can also be combined with the orthogonal frequency division multiplexing (OFDM) [15] similar to the MIMO-OFDM system.

Due to the severe crosstalk in multipath environments, it is usually believed that OAM multiplexing is only useful in pure LOS environment. Nevertheless, due to misalignment and imperfect and finite-sized designs, crosstalk exists even in the absence of multipath. The crosstalk of conventional OAM multiplexing is typically around -12 dB or worse at higher-order OAM mode. Due to the (crosstalk) interference, increasing the transmitted power will increase the interference. Thus, the signal-to-interference-plus-noise ratio (SINR), or the effective signal-to-noise ratio (SNR), of a communication link cannot increase unlimitedly by simply increasing the transmitted power. Therefore, the modulation is usually restricted to 16 quadrature amplitude modulation (16-QAM) or less in experimental demonstrations of OAM communications [1], [6], [11].

Unlike these OAM-related works in literature, we study the OAM multiplexing in rich multipath (highly reverberant) environments (which has not been investigated yet). Specifically, OAM waves carrying independent data are sent out from the transmitter. The receiver (equipped with multiple antennas) uses spatial equalizer for data detection. With the spatial equalization (e.g., zero-forcing equalizer [16]), the crosstalk can be eliminated. Hence, the effective SNR of the communication link increases with increasing transmitted power and modulation order higher than 16-QAM (e.g., 64-QAM) can be used, enabling higher data rate.

## II. MEASUREMENTS

As mentioned before, OAM waves can be generated using various methods. In this work, we employ geometric-phase based in-house OAM metasurfaces [17] instead. Figure 1 shows the measurement setup of a 2×2 OAM-based MIMO system in a reverberation chamber (RC). OAM waves are generated by placing two OAM metasurfaces (with modes $l = 1$ and $l = 2$) on top of two identical horn antennas. A 10-cm foam support is inserted in between the metasurface and the horn aperture. The horn antennas are beneath the foam supports and only partially shown in the photo. In conventional LOS

Manuscript received xx, 2019. This work was supported by the National Natural Science Foundation of China under Grant 61801366.

X. Chen, W. Xue, H. Shi are with the Faculty of Electronic and Information Engineering, Xi'an Jiaotong University, Xi'an 710049, China.
J. Yi is with the Key Laboratory of Integrated Services Networks, Xidian University, Xi'an 710071, China.
W.E.I. Sha is with the Key Laboratory of Micro-Nano Electronic Devices and Smart Systems of Zhejiang Province, College of Information Science and Electronic Engineering, Zhejiang University, Hangzhou 310027, China.

scenario, OAM antennas should be adopted at the receiver to detect the OAM waves. However, in highly reverberant environments (such as the RC), the OAM orthogonality will be destroyed completely during propagation. Instead, two (non-OAM) slot antennas are used as the receiving antennas. The slot antennas were designed for mobile phone applications [18], covering the bandwidth of the OAM metasurface. Note that there are actually eight slot antennas on the mobile chassis. In order to avoid strong crosstalk between the receiving antennas, we choose two farthest separated slot antennas with lowest mutual coupling and fixed them for the entire RC measurements. So, we have a 2×2 MIMO system.

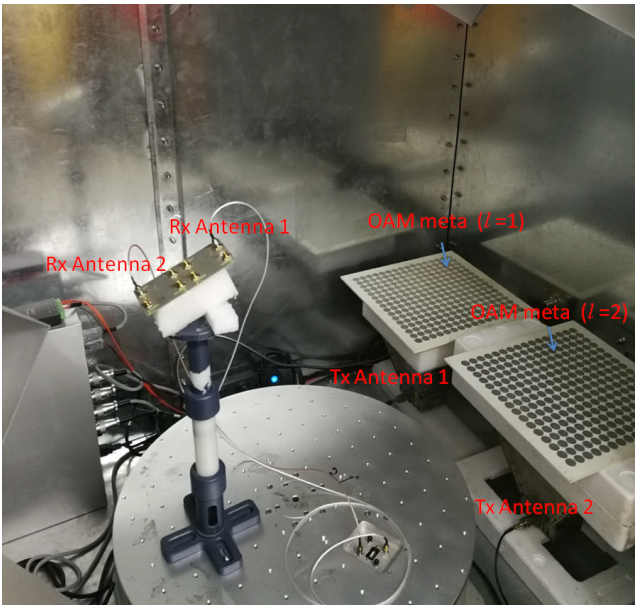

Fig. 1. Photo of measurement setup of OAM-based 2×2 MIMO system in RC.

The RC is an electrically-large metal cavity emulating a rich multipath environment [19]. The used RC in this work has a size of 1.50 m × 0.92 m × 1.44 m. There are one turn-table platform and two mechanical mode stirrers (one horizontal and one vertical) inside the RC. The horn antennas (below the OAM metasurfaces) are pointing towards the horizontal stirrer to reduce the unstirred field. During the measurement, the platform rotates stepwise to 20 angles with an angular step of 18º. At each platform state, the two stirrers rotate stepwise (and simultaneously) to 20 angles with an angular step of 18º, resulting in 400 samples (20 platform-stirring samples × 20 mode-stirring samples). At each stirring state, the channel transfer function is sampled from 5 to 5.2 GHz with a frequency step of 1 MHz using a vector network analyzer (VNA). Since only a two-port VNA is available, we have to repeat the measurement procedure four times (for the four subchannels of the 2×2 MIMO system). Each time the platform and stirrers automatically move back to their initial states so that the environments for the four subchannel measurements are basically the same. The MIMO channel measurement conducted in this way is equivalent to the switch-based MIMO channel sounding (in controllable environments) [20].

In order to compare the OAM multiplexing with the conventional MIMO system, we repeat the whole measurement procedure by removing the OAM metasurfaces from the horn apertures. In this case, we have a conventional 2×2 MIMO system with two horn antennas at the transmitter and two slot antennas at the receiver. (For the conciseness, the photo of the measurement setup of this case is not shown.) We refer to this case as MIMO system without OAM waves and the previous case where OAM metasurfaces are placed on top of the horn antennas as MIMO system with OAM waves. We will compare the performances of the two MIMO systems with and without OAM waves in the next section.

### III. RESULTS AND DISCUSSIONS

The ergodic capacity of a 2×2 MIMO system can be calculated as [21]

$$C(f) = \mathrm{E}\left\{\log_2\left[\det\left(\mathbf{I}+\frac{\gamma}{2}\mathbf{H}(f)(\mathbf{H}(f))^{\dagger}\right)\right]\right\} \quad (1)$$

where $\mathbf{H}(f)$ denotes the 2×2 MIMO channel transfer function at frequency $f$, the superscript $^{\dagger}$ denotes conjugate transpose, $\mathbf{I}$ is the identity matrix, $\gamma$ denotes the reference signal-to-noise-ratio (SNR), det is the determinant operator, log is the logarithm operator, and E denotes the expectation (over the random channel $\mathbf{H}$). In the calculation of the ergodic capacity, the expectation is approximated by averaging over the 400 samples (cf. Section II). In a well-stirred RC, the entries in $\mathbf{H}$ follow zero-mean Gaussian distribution (i.e., Rayleigh fading channel). Note that $\mathbf{H}$ in (1) is the measured channel transfer function matrix normalized by the square root of the average power transfer function of a reference measurement (where the efficiency of the reference antenna is known and calibrated out) [22]. Also note that (1) holds for both MIMO systems with and without OAM waves and we use the same reference measurement for both cases.

Figure 2 shows the ergodic capacities of the two MIMO systems with and without OAM waves at an arbitrary reference SNR, i.e., $\gamma = 15$ dB. Note that OAM metasurfaces have about 2-dB insertion losses. In order to focus on the effects of OAM waves, we measured the insertion losses of the meatasurfaces (about 2 dB) separately and calibrated them out. As can be seen, the capacities of the MIMO systems with and without OAM waves are approximately at the same level over the operating bandwidth of the OAM metasurfaces. The results may be surprising. Given the fact that the OAM orthogonality is destroyed in the highly reverberant environment, one may think the capacity of the OAM multiplexing would be much lower than that of the conventional MIMO system in the RC. Nevertheless, even though the phase fronts of the OAM waves are distorted in the RC, the information carried by the original OAM waves will not be lost. The transmitted waves from two horn apertures can be regarded as a superposition of rays and they will constructively and destructively interfere with each other through multiple reflections off metal walls and mode stirrers. This transmitted process can be mathematically expressed as a random matrix $\mathbf{H}$. The insertion of the metasurface is nothing but change the directions and redistribute the complex amplitudes of rays, expressed as a certain matrix **M**. The matrix **M** should be block-diagonal and unitary if ignoring the imperfect designs (i.e., Jones matrix for each metasurface element is unitary) [9]. Thus, it is easy to prove that $E\{\mathbf{HH}^\dagger\}=E\{\mathbf{HM}(\mathbf{HM})^\dagger\}$ [9]. The above is ray physics for understanding the comparable channel capacity. As described above, the received signal at each slot antenna is a mix of the two transmitted signals, i.e., there are strong crosstalk. The crosstalk can be readily mitigated using the spatial equalizer to (approximately) diagonalize the channel matrix **H** (in the data domain) for data detection [16]. The small discrepancies between the two capacities are due to the frequency dependence of the OAM waves generated by the imperfect OAM metasurfaces.

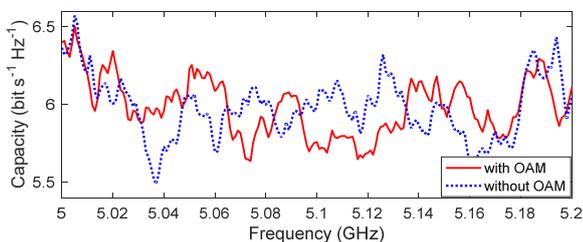

Fig. 2. Ergodic capacities of the 2×2 MIMO systems with and without OAM waves at 15-dB reference SNR.

In order to have a closer look at the OAM effects on multiplexing, we assume MIMO-OFDM systems with and without OAM waves and compare their performances of bit error rate (BER). Specifically, we assume the systems operate over the entire 200-MHz bandwidth (i.e. 5-5.2 GHz). The coherence bandwidth [19] (with 0.5 correlation threshold) of the channel in the RC (for the measurement setup) is about 2 MHz. To cope with the frequency selectivity of the channel, the number of OFDM subcarriers must be larger than 100 (200 MHz / 2 MHz). To be safe, we assume 512 subcarriers in this work. The length of the cyclic prefix is set to 128 (i.e., 640 ns) that is sufficiently larger than the excess delay spread of the channel. Each of the subcarriers is loaded with a 64-QAM symbol. The transmitter sends two independent data streams. The simulated streams are convolved with the MIMO channel impulse responses (obtained by applying the inverse Fourier transform to the measured channel transfer functions). At the receiver, we use the zero-forcing equalizer [16] (to remove the interference) for data detection. We repeat the simulation procedure for MIMO-OFDM systems with and without OAM waves. Figure 3 shows the raw (uncoded) BER performances for two cases. (The BER is calculated as the ratio of the number of falsely detected bits to that of the totally transmitted bits. It is calculated for each channel sample and then averaged over all the 400 channel samples.) As can be seen, the OAM multiplexing system has the same performance as the conventional spatial multiplexing system. Note that the BER performance is calculated over the entire bandwidth. Hence, we do not see the frequency variation as in the capacity plot (cf. Fig. 2). Since we use a 2×2 MIMO system to transmit two data streams, there is no spatial diversity (i.e., the diversity order is one). Thus, at high SNR, the BER is around $K\gamma^{-1}$ [16], where $K$ is a constant and $\gamma$ denotes the SNR. The BER limit is plotted as a reference in the same figure. As can be seen, the simulated BER follows tightly with the BER limit at high SNR (> 20 dB).

Figure 4 shows the correlations between the two transmit antennas with and without OAM waves. As can be seen, the correlations are frequency dependent. As can be seen by comparing Figs. 2 and 4, generally speaking, a higher correlation corresponds to a lower capacity in the frequency domain (even though the correlations are fairly small). Nevertheless, the correlations averaged over frequency are about the same, which explains the identical BER performance. Also note that the spatial channel in the RC follows the Kronecker structure, i.e., correlations at the receive side are independent on that at the transmit side [19], and the correlations at the receive side are smaller than 0.1 for both cases (due to rich scattering process of waves). For the conciseness of the letter, they are not shown here. Unlike the conventional OAM multiplexing where data detection solely relies on the orthogonality between the OAM modes and no higher than 16-QAM can be used (due to the intrinsic crosstalk which limits the effective SNR), it is demonstrated here that 64-QAM can be supported by combining OAM transmission with zero-forcing equalization because the zero-forcing equalizer can remove the interferences from both multipath and intrinsic crosstalk. Nevertheless, it should be noted that the superior performance of the spatial equalizer to the conventional (discrete) Fourier transform based OAM detection is at the cost of increased computational complexity, especially as the number of data streams increases.

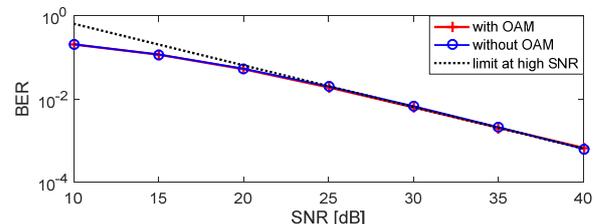

Fig. 3. Uncoded BER performances of the 2×2 MIMO-OFDM systems with and without OAM waves.

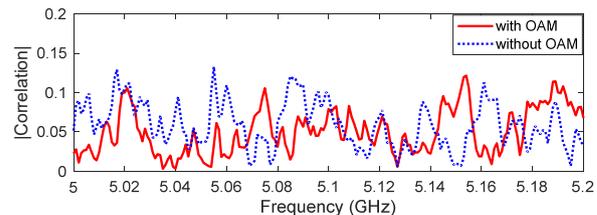

Fig. 4. Correlation magnitudes between the transmitted signals with and without OAM waves.

## IV. Conclusion

Unlike the previous works, this letter studies OAM multiplexing in a highly reverberant environment. It was shown that OAM multiplexing combined with zero-forcing equalizer and OFDM could not only cope with the rich multipath effects but also support high-order modulation (e.g. 64-QAM as opposed to the previous 16-QAM) transmission. And the OAM multiplexing system exhibited comparable performance to the conventional MIMO system. Thus, OAM communication should not be confined only to LOS environments.